\documentstyle[art12,bezier]{article}

\begin{document}

\renewcommand{\thefootnote}{\fnsymbol{footnote}}

\thispagestyle{empty}

\hfill \parbox{45mm}{
{\sc CTP\ \#2253} \par
November 1993}

\vspace*{15mm}

\begin{center}
{\LARGE On the Absence of Localized Curvature \par
        in the Weak Coupling Phase of Quantum Gravity.}

\vspace{22mm}

{\large Giovanni Modanese}%
\footnote{This work is
supported in part by funds provided by the U.S. Department of Energy
(D.O.E.) under contract \#DE-AC02-76ER03069.}

\medskip

{\em Center for Theoretical Physics \par
Laboratory for Nuclear Science \par
Department of Physics \par
Massachusetts Institute of Technology \par
Cambridge, Massachusetts, 02139, U.S.A.}

\bigskip

\bigskip

Submitted to: {\it Phys.\ Lett.\ B}

\end{center}

\vspace*{20mm}

\renewcommand{\thefootnote}{\arabic{footnote}}
\setcounter{footnote} 0
\begin{abstract}
In the weak field expansion of euclidean quantum gravity, an analysis of the
Wilson loops in terms of the gauge group, $SO(4)$, shows that the correspondent
statistical system does not develope any configuration with localized curvature
at low temperature. Such a ``non-local'' behavior contrasts strongly with that
of usual gauge fields. We point out a possible relation between this result and
those of the numerical simulations of quantum Regge Calculus. We also give a
general quantum formula for the static potential energy of the gravitational
interaction of two masses, which is compatible with the mentioned vacuum
structure.

\bigskip \bigskip

\end{abstract}

\newcommand{\beq}{\begin{equation}}    \newcommand{\la}{\langle}
\newcommand{\eeq}{\end{equation}}      \newcommand{\ra}{\rangle}
\newcommand{\beqa}{\begin{eqnarray}}   \newcommand{\pa}{\partial}
\newcommand{\eeqa}{\end{eqnarray}}     \newcommand{\half}{\frac{1}{2}}

\newcommand{\m}{\medskip}

\newpage

In the familiar quantum gauge theories, employed to describe strong and
electroweak interactions, the Wilson loops
\beq
  W(C) = \la {\rm Tr} \, {\rm P} \, \exp \oint_C dx^\mu
  A_\mu(x) \ra_0
\eeq
are the most relevant and physically meaningful invariants. It is known that in
the euclidean theory the static potential energy $U(L)$ of two sources of the
gauge field is related to a loop of temporal size $T$ and spatial size $L \ll
T$ by the formula
\beq
  e^{- \hbar^{-1} T U(L)} = W(L,T) .
\eeq
In the strong coupling limit this expression leads to the confining potential
$U(L)=kL$, while in the weak coupling limit one easily recovers the Coulomb
potential $U(L)=-e^2/L$.

\m
It may thus be interesting to test the behaviour of $W$ in perturbative quantum
gravity. As we shall see, although quantum gravity is not a full gauge theory
in the usual sense, the loops are geometrically well defined and contain useful
information about the structure of the vacuum state.

The field which enters the loops in this case is the Christoffel connection
$\Gamma^\alpha_{\mu \beta}$, or the spin connection $\Gamma^a_{\mu b}$
\cite{hehl}. By definition, the (classical) loop ${\cal W}(C)$ is the trace of
the matrix ${\cal U}(C)$ which performs the parallel transport of vectors along
a closed curve $C$
\beq
  {\cal W}(C) = {\rm Tr} \, {\cal U}(C) .
\label{klu}
\eeq
The ``holonomic'' and ``anholonomic'' components of ${\cal U}$ are given
respectively by
\beq
  {\cal U}^\alpha_\beta(C) = {\rm P} \, \exp \oint_C dx^\mu
  \Gamma^\alpha_{\mu \beta}(x)
\eeq
and
\beq
  {\cal U}^a_b(C) = {\rm P} \, \exp \oint_C dx^\mu
  \Gamma^a_{\mu b}(x) .
\eeq
It can easily be shown \cite{wilson} that ${\cal W}(C)$ is invariant with
respect to coordinates transformation and also, as a functional
 of $\Gamma^a_{\mu
b}$, with respect to local Lorentz transformations of the vierbein.

\m
We recall that Einstein's gravity, written in the vierbein formalism, is a
gauge theory of the Lorentz group (i.e., the action is invariant under local
Lorentz transformations of the vierbein), but not of the whole Poincar\'e group
$ISO(3,1)$. A complete gauge formulation can be obtained only introducing some
auxiliary fields $q^a$ \cite{grinar}. So it is not possible to consider in
(3+1) dimensions, like in (2+1)-gravity \cite{witten}, the holonomies of the
Lie-algebra-valued connection
\beq
  {\cal A}_\mu(x) = e^a_\mu(x) P_a + \Gamma^{ab}_\mu(x) \omega_{ab} ,
\eeq
where $P_a$ and $\omega_{ab}$ are the generators of translations and of the
Lorentz transformations.

\m
The geometrical meaning of ${\cal W}(C)$ in euclidean quantum gravity is the
following. During the parallel transport of a vector $V$, its length, given by
\beq
  |V|^2 = V^a V^b \delta_{ab} = V^\mu V^\nu g_{\mu \nu}(x) ,
\eeq
does not change. If we transport $V$ along a closed curve $C$, returning to the
starting point, we obtain another vector $V'$, which has the same length of
$V$, and differs from it only in the orientation. Hence we have for any vector
and any curve
\beq
  V^a V^b \delta_{ab} = {V'\,}^a {V'\,}^b \delta_{ab} =
  {\cal U}^a_c(C) V^c \, {\cal U}^b_d(C) V^d \, \delta_{ab} ,
\eeq
or, in matrix notation
\beq
  {\cal U}^T(C) \, {\cal U}(C) = {\bf 1} .
\eeq
The matrix ${\cal U}$ belongs then to $SO(4)=[SO(3)]_1 \times [SO(3)]_2$ and it
is known \cite{wybour} that for weak fields its trace has the form
\beq
  {\cal W}(C) = 4 - (\theta_1^2 + \theta_2^2) ,
\eeq
where $\theta_1$ and $\theta_2$ are the two independent angles which describe
the $SO(4)$ rotation of $V$.

\m
Let us now evaluate perturbatively the quantum average of ${\cal W}(C)$.
Following the usual approach, we decompose the metric $g_{\mu \nu}(x)$ as
\beq
  g_{\mu \nu}(x) = \delta_{\mu \nu} + \kappa h_{\mu \nu}(x); \qquad
  \kappa = \sqrt{16\pi G} ,
\eeq
and we interpret $\delta_{\mu \nu}$ as a classical background, while $\kappa
h_{\mu \nu}(x)$ is regarded as a small quantized perturbation. The
Einstein-Hilbert lagrangian is then splitted into a quadratic part and into
interaction vertices; the first two vertices, respectively proportional to
$\kappa$ and $\kappa^2$, connect 3 and 4 fields $h$. Let us denote by $W^{(n)}$
the $n$-th term in the expansion of the P-exponential defining $W$. The leading
contribution to $W$, of order $\hbar \kappa^2$, is given by $W^{(2)}$ with one
bare propagator, namely
\beq
  W^{(2)} = \oint_C dx^\mu \oint_C dy^\nu
  \la \Gamma^\beta_{\mu \alpha}(x) \, \Gamma^\alpha_{\nu \beta}(y) \ra .
\label{dle}
\eeq

The following two contributions to $W$, of order $\hbar^2 \kappa^4$, are given
by the term $W^{(4)}$ with two bare propagators and by the term $W^{(3)}$ with
three propagators and one $\kappa$-vertex. Finally, the three contributions of
order $\hbar^3 \kappa^6$ are given by the term $W^{(6)}$ with three
propagators, by the term $W^{(4)}$ with four propagators and one
$\kappa^2$-vertex and by the term $W^{(2)}$ with the radiatively corrected
propagator.

\m
What is remarkable, and quite easily shown \cite{gtrips}, is that the leading
term
(\ref{dle}), of order $\hbar$, vanishes in Einstein's theory. As a matter of
fact, it can be proved \cite{wilson} that this vanishing does not strictly
depend on the dynamical content of Einstein's action, but is only due to the
symmetries of the propagator, to the Poincar\'e invariance of the background
and to the absence of a dimensional coupling (apart from the overall factor
$\kappa^{-2}$) in the action. In particular, these properties are
maintained in theories of euclidean quantum gravity with an ``improved''
analytic continuation procedure \cite{greens}.

{}From the geometrical point of view, the vanishing of $W$ to order $\hbar$
means
that, in the same approximation,
\beq
  \la \theta_1^2 + \theta_2^2 \ra_0 = 0 .
\eeq
If the variance of the angles $\theta_1$ and $\theta_2$ is zero, the angles
themselves have to vanish identically in any configuration, that is
\beq
  {\cal U}(C) = {\bf 1} \ \ {\rm for \ any} \ C.
\eeq
This is a very strong geometrical statement, as it implies that, still
to order $\hbar$, all the weak field configurations which effectively enter
the functional integral
\beq
  z = \int d[h] \, \exp \left\{ - \hbar^{-1} \, S[h] \right\}
\label{dks}
\eeq
have no curvature. In other words, the curved configurations -- which possibly
dominate in other regimes -- are in this approximation totally suppressed.

\m
This unexpected situation should be compared with what happens, for
instance, in a ordinary $SO(N)$ gauge theory. In this case we have
\beq
  W(C) = N - \la \theta_1^2 + \theta_2^2 + ... + \theta_{N-2}^2 \ra_0
\eeq
and the perturbative computation shows that the
leading term $W^{(2)}$ does not vanish. So the matrices of the parallel
transport in the ``internal'' gauge manifold, considered configuration by
configuration, are not equal to the identity matrix. Interpreting $\hbar$ as
the temperature $\Theta$ of an equivalent statistical system, we see that when
$\Theta$ grows from zero to some small value -- such that we may disregard
$\Theta^2$ or higher orders -- the Yang-Mills fields develop ``localized
excitations'', i.e.\ regions of various sizes where the Yang-Mills curvature is
not vanishing.
All this does not happen for the gravitational field, which, to say so,
does not ``boil'', but remains essentially in a flat state.

This picture also explains the fact \cite{correl} that to leading order
in $G$ all the invariant correlation functions of the Riemann curvature
are vanishing: this is a natural consequence of the absence of localized
excitations.

Moreover, it is very interesting to make a comparison with the results,
in the same regime of small $G$, of the Montecarlo simulations of
discretized quantum gravity. The classical simulations of Hamber
\cite{Hamber}, based on Regge Calculus, show that there is a transition,
in the phase space of quantum gravity, between a ``smooth'' phase
(for large values of $G$) and a ``rough'' phase (for small values of $G$).
The nature of the transition, the critical indices and the possibility
of a continuum limit have been investigated too. The rough phase is
believed to be physically unacceptable, since in this phase the simplices,
which are the basis of the spacetime lattice, are collapsed and
the fractal dimension is much less than four.

Our results give a possible justification of this phenomenon.
In fact, the absence of localized curvature for continuum quantum gravity
in the weak coupling limit suggests that the quantum Regge Calculus,
which is a discretization scheme
based on the curvature, is not really a description of lattice quantum
gravity in this limit.

\m
It can also be shown \cite{wilson} that the introduction of a small external
source in the functional integral (\ref{dks}), breaking the Poincar\'e
invariance of the background, gives rise in general to a non vanishing
contribution to the loop proportional to $\hbar$. In this case we may have
excitations with localized curvature, but they are very small, since their
variance is proportional to $\kappa^3$ instead of $\kappa^2$ (they are in fact
originated by the non-linear interaction of gravitons).

\m
The vanishing of $W$ to leading order in gravity raises the problem of
finding another invariant expectation value of the quantized field which gives
the static potential energy between two masses. This problem has a well defined
solution indeed \cite{energy}. One starts from the following known formula of
euclidean field theory
\beq
  {\cal E} = \lim_{T \to \infty} - \frac{1}{\hbar T} \log
  \frac{\int d[\phi] \, \exp \{ - \hbar^{-1} ( S_0[\phi] +
  S_{\rm Inter.}[\phi,J] ) \} }
  {\int d[\phi] \, \exp \{ - \hbar^{-1} S_0[\phi] \} },
\eeq
where $\phi$ is a quantum field and $J$ is a classical source coupled to
$\phi$,
which is switched off outside the interval $(-\half T,\, \half T)$. ${\cal E}$
represents the energy of the ground state of the system.
We shall only show here how this formula works in the case of
a weak gravitational field on a flat background.
Replacing $\phi$ with
the gravitational field, $S_0$ with Einstein's action and $J$ with two
particles of masses $m_1$, $m_2$, following the trajectories
\beq
  x^\mu(t_1) = \left( t_1,\, -\frac{L}{2},\, 0,\, 0 \right); \qquad
  y^\mu(t_2) = \left( t_2,\, \frac{L}{2},\, 0,\, 0 \right) ,
\label{lhj}
\eeq
we find (for a weak field)
\begin{displaymath}
  {\cal E} = \lim_{T \to \infty} - \frac{1}{\hbar T} \times
  \hskip 12 truecm
\end{displaymath}
\beq
  \ \times \log
  \frac{\int d[h] \, \exp \hbar^{-1} \left\{ - S_{\rm Einst.}[h] -
  m_1 \int dt_1 \, \sqrt{1-h_{00}[x(t_1)]} -
  m_2 \int dt_2 \, \sqrt{1-h_{00}[y(t_2)]} \right\}}
  {\int d[h] \, \exp \{ - \hbar^{-1} S_{\rm Einst.}[h] \} } .
\label{per}
\eeq
It is easy to verify that to leading perturbative order this gives
the correct result
\beq
  {\cal E} = m_1 + m_2 - \frac{m_1 m_2 G}{L} .
\eeq

The geometrical content of eq.\ (\ref{per}) and of the general expression for
${\cal E}$ \cite{energy} is compatible with the absence of localized
excitations. In fact, the interaction energy arises because there
is a difference in the expectation value of the {\em total} proper time
measured
along the trajectories (\ref{lhj}) of the two particles and along that of their
center of mass. Such a ``non-localized''
feature of the interaction energy is also in agreement
with the known fact that it is impossible in General Relativity to localize
covariantly the energy of the gravitational field.

\m
Eq.\ (\ref{per}), like the corresponding ones in QED or QCD, has the physically
appealing feature of showing how the force between the sources ultimately
arises from the exchange of massless bosons. However, let us make a closer
comparison with electrodynamics. In that case the analogue of the functional
integral which appears in the logarithm of (\ref{per}) has the form
\cite{fischl}
\beq
  \left< \exp \left\{ e \int_{-\frac{T}{2}}^{\frac{T}{2}} dt_1
  A_0[x(t_1)] - e \int_{-\frac{T}{2}}^{\frac{T}{2}} dt_2
  A_0[y(t_2)] \right\} \right> .
\label{vge}
\eeq
(The two charges have been chosen to be opposite: $q_1=e$, $q_2=-e$.) Reversing
the direction of integration in the second integral and closing the contour at
infinity, one is able to show that the quantity (\ref{per}) coincides with the
Wilson loop of a single charge $g$, thus giving a gauge invariant expression
for the potential energy.

In gravity this is not possible: we may imagine that an expression like
(\ref{per}) could be obtained in the first-order formalism (with $A_0$ replaced
by the tetrad $e^0_0$), but the masses necessarily have the same sign,
so the loop cannot be closed.

\m
In conclusion, we believe that the above results, due to their generality
(within the approximation we considered) give a new insight into some physical
properties of the quantized gravitational field at energies which are small
with respect to the Planck scale. These properties are true either if we regard
quantum gravity as a (not yet established) fundamental theory, or as an
effective quantum field theory which has General Relativity as its classical
limit and goes to some more fundamental theory at very short distances.

\m
I would like to thank R.\ Jackiw for his kind hospitality at M.I.T.\ and for
very helpful discussions. I also have benefited from useful discussions
with D.\ Cangemi, D.Z.\ Freedman and C.\ Lucchesi and from an earlier
conversation with M.\ Toller. Financial support was provided to the
author by the ``A.\ Della Riccia'' Foundation of Florence, Italy.

\end{document}